\documentstyle[12pt,a4]{article}

\begin{document}
\title{Characterisation of Fourteenth-Century Bell-Casting Pit in  Old Town
Hall Sibiu, Romania}
\author{Agata Olariu\\
{\em National Institute for Physics and Nuclear Engineering}\\
{\em P.O.Box MG-6, 76900 Bucharest Magurele, Romania}\\
Petre Munteanu-Besliu\\
{\em Br\"{u}kenthal Museum, Sibiu, Romania}}

\maketitle

\section{Introduction}

During the Middle Ages the skill of moulding bells was extremely valuable. The
craftsmen smelting bells also produced cannons and baptismal fonts 
(Besliu 1989).
In the long run the bells were melted when cannons were needed and  these
were transformed again when the cannons were no longer necessary. For this 
reason one
preserves only a few old bells. An 1507' bell is used at the Hospital
Church in Sibiu. Another bell, dated 1626, is present in the collection of the
History Museum from Sibiu, and others bells, remelted several times are in the
tower of the Evangelic Church from Sibiu.
Installations for moulding or fragments of loam patterns obtained 
as a result of 
the moulding of bells are even more rare. Some oven installations have been
discovered in Winchester (UK), Leibgott (Denmark), Visegrad (Hungary) 
(Russel 1990).
The similar discovery from the Old Town Hall from Sibiu, Romania can be linked
with the above mentioned examples but also with other uncoveries in the 
neighborhood: a wooden building from the XIII$^{th}$-XIV$^{th}$ centuries, 
a crematory for garbage, and the foundations of the 
previous buildings, joined after 1470 to form the Old Town Hall. \\
The installation for smelting copper-alloy discovered in Sibiu in the big
inner yard of the Old Town Hall was situated in a pit located near the 
parochial church. 
 The gothic church was build in the middle of XIV$^{th}$ 
century and was enlarged at the beginning of XV$^{th}$ century. The two 
periods
and the phases of the construction are linked with the casting of a bell
(Roth 1908).
The installation was composed of two loam
structures. The first one, a hearth on the ground pit 
at -2.70 m appeared as a  circular structure with 1.40 diameter,  
as shown in Figure 1. The northern 
part was enlarged.
The loam structure is 0.24 m
tall and has a narrow ditch in the lower part. Near this structure were
uncovered some pieces with an aspect of  copper-alloy. The second 
loam structure has an ogival form with the raised brims  and
the surface slightly inclined with 3-4$^{\circ}$. Some pieces of metal and
slag of copper-alloy have been also found
on its surface. Between these two hearths there is a slag layer.
The two hearths belonged to a furnace for melting copper-alloy. This
furnace is unique but the other archaeological uncoveries in Europe
are some pits with stone furnace on which the mould was fixed (Russel 1990).\\
In this work we have made an analytical investigation by neutron activation
analysis and X-ray fluorescence of a number of pieces found on the two
hearths and the slag layer discovered at
the Old Town Hall
from Sibiu, Romania and some bells from the surrounding region of Sibiu to answer
the question if there is a relationship between these two hearths and 
the slag layer. We also examined the possibility of preparing  of the alloy 
for
moulding of a bell in the furnace discovered at the Old Town Hall.

\section{                         Experimental Analysis}

The objects which have been analysed are listed in Table 1. 
The samples of metal and slag presented in Table 1 have been analysed by means 
of the X rays fluorescence (XRF) and neutron activation analysis (NAA) 
methods.
XRF has been used for a qualitative analysis of the 
samples and for the quantitative determination of lead, which cannot be 
observed by NAA.
The objects have been cut with a hard steel knife in order to
 obtain relatively homogeneous
samples. The sample Ie, a piece with a surface of 3 cm$^2$ from level I of 
the oven has been
specially prepared 
(by extracting an evident piece of white metal which was cut and polished) for
the determination of concentration of metal by XRF.
The bells have been
sampled by drilling with discarding of the surface area to avoid the corrosion
products and other contamination. The resulting quantity of powder was so 
small
that it did not permit the analysis of lead content by XRF, so that only 
the analysis by neutron activation was possible. 
The XRF analysis  has been performed with the aid of a triple source of 
Pu$^{238}$, of 
3x33 mCi. The fluorescence X rays have been detected with a GeHP detector and 
the characteristic spectra have been counted on a  computer with a 
multi-channel analyzer  interface.
The following elements have bean detected: Cu, Fe, Pb and Sn.\\
As concerning neutron activation analysis, samples of approx. 10 mg have been 
cut  with  a   hard  steel  knife  with  care  to  obtain   relatively 
homogeneous samples.
These samples have been introduced in 
polyethylene foils and have been 
irradiated at VVR-S Reactor of National Institute of Physics and Nuclear 
Engineering Bucharest-Magurele, at a rabbit system at a neutron flux of 
2.5$\cdot$ 10$^{12}$ neutrons/cm$^{2}
\cdot$ s, for several periods of time: 20 s, 2 min and 30 min. The induced 
$\gamma$ 
radioactivity in the samples has been detected by a 135 cm$^{3}$ GeLi detector
(EG \& G Ortec) with 2.7 keV resolution. The samples have been counted after 
a cooling time of 2 minutes, and again after 1 day, 3 days and 10 days. 
14 elements have been noticed: Ag, Al, As, Au, Cu, Ca, Fe, K, Mg, Mn, Na, Sb, 
Sn and Sc.

\section{                    Results and Discussions}

In Table 2 is presented the picture of concentrations of elements detected in 
samples of the two hearths and the slag layer by the NAA, and the lead 
content from XRF analysis. The Table 3 shows the results of NAA for the bell 
samples. The results are expressed
in parts per million (ppm) and when the result was greater than 10000 ppm it 
was
given in \%. The relative errors of measurements are $<$10\% 
for both NAA and Pb content.\\
As concerning the analyses of hearths samples 
given in Table 2,
the samples contain tin in  relatively high 
concentrations [2\% - 37\%]. The sample Ie, Hearth level I has the highest 
content of tin: 36.7\% .
We mention that tin as impurity is rarely found in
copper ores (Craddock 1981) and its presence in high concentrations alongside 
with the
copper can be explained only by a deliberate action of alloying.
The samples of the Hearth level II no. IIas, IIbs, IIcs and IIa, IIb, IIc,
surface and deepness, have a lower content of  copper
and tin, but the ratio Sn/Cu (Figure 2) is however in the range of 
values of this ratio
for the samples no. Ia, Ib, Ic, Id of Hearth level I as well as for the 
samples
no. Za, Zb, Zc, Zd for the
slag. This fact can be explained by a relationship 
between Hearth level I, Hearth level II and the Slag layer. 
Iron, in high concentrations in samples of Hearth level II can be 
explained by the fact that iron oxides were deliberately added in melted 
charge, 
these acting as fondant. The added iron can be found in slag.
From the viewpoint of iron, samples IIas and IIbs
are very different from the others: the ratio Fe/Cu is very high, approx. 15
(Figure 3).
The samples Z found in the intermediate layer between the two hearths, known 
as
slag samples, have a concentration of Cu in the range 25-36\%, which is
high compared to the usual concentration of Cu in slag of about 4-5\% .
In Figure 4 the diagram of the ratio Pb/Cu versus the ratio Sn/Cu
shows a close relationship between the hearth samples I and II and the slag 
layer, from the point of view of elements: Cu, Sn and Pb.
We remark a similar composition of samples Z, slag with the Hearth level
 II, 
both from the point of view of Fe and the impurities Al, Na, K,
Sb and Mn.  
The content of tin and copper and the concentration of the other minor 
elements
suggest that the two hearths level I and level II belonged to a furnace for
alloying of copper and tin to obtain specific objects like bells, statuar art
objects or weapons.
The sample of ash, C2,  approaches the  samples from the two hearths and slag.
The content of Pb presents a background of concentrations with values not 
very straggled with a signal 
of C$_{Pb}$=21.8\% from the sample C1. This sample is very different in
 comparison with
the other samples.
The significance of sample C1, piece of metal-cake, remains a question mark 
due to its high
content of lead and silver and the lack of tin; it may suggest the base 
mineral
for producing the silver: galena and by inference it may suggest the presence 
of a mint. 
Initially it was believed that in this furnace was prepared the alloy for the 
coins but the composition of some binary denars (Besliu 1987) , (Table 4), 
silver and copper, with high concentrations of 
silver, and the presence of silver only in traces in the samples from the 
2 hearths eliminates this assumption.
Also a NAA analysis of some coins (Besliu 1987) point out the 
lack of the element Sn, 
considered a very important presence in the hearth samples.
The concentration of zinc for the samples from Table 2 was for the 
given experimental conditions $<$1000 ppm.
The analyses of the two hearths discovered 
at Old Town Sibiu led to  the further analysis of some  tin-alloy objects, 
like
bells from the neighborhood of Sibiu. The results of the analyses of 4
Transylvanian bells
shown in Table 3 indicate the similarity of their elemental composition  
and also the similarity with the samples from the Hearth level I, considered 
as
representative for the furnace. The diagram 
from 
Figure 5, the ratio (Au*100+Ag*10+Sb+As)/Cu versus the ratio Sn/Cu, shows that
 the 
concentration for the samples of Hearth level I
agrees very well with the 4 bells. The B1, B2 and B3 bells from the Sibiu 
region and
Middle Ages have a close pattern
composition and the B4 bell, XIX century has a somewhat a relative different 
composition. (Figures
5 and 6). 
This suggests the possibility that in these hearths was prepared a specific
copper-tin alloy for smelting bells. 
Also,  in the Table 5 it is shown the composition of 2 famous bells 
(Hanson 1978): Whitechapel bell and Christ Church bell, made in the same 
foundry in London.
For comparison, in Figure 5 it is represented the composed ratio
(Au*100+Ag*10+Sb+As)/Cu versus the ratio of concentrations Sn/Cu for both
Transylvanian bells and the London-made bells. Also in Figure 6 it is 
represented the composed ratio (Au*100+Ag*10+Sb+As+Sn)/Cu versus the ratio 
Fe/Cu for the same bells. Figures 5 and 6 can interpreted that the 
craftsmen for Transylvanian bells had a
smelting formula which is somewhat different of that of London-made bells.

\section{                              Conclusions}

The  elemental analyses from this study suggest a relationship between
Hearth level I, Layer of Slag and  Hearth level II and the fact that the
 hearths
preserve traces of an activity of alloying of copper with tin. Also one 
observes
that  the elemental composition of the alloy.
for the 4 bells from the Sibiu region analysed in this study
is very similar with that of the metallic pieces found at the two hearths,
discovered at Old Town Hall, Sibiu. We can conclude that at these hearths was
prepared a copper-tin alloy for casting bells with the same formula with that
of analysed Transylvanian bells.\\
\newpage
{\large \bf References}\\
\noindent
Besliu P. und N\"{a}gler T., 1989, Die Arch\"{a}ologischen Granges im
Hermannst\"{a}dter\\
\hspace*{1cm}Alten Rathaus, {\it Forschungen Zur Volks und Landeskunde},
{\bf 32} (2), 29-40\\
Besliu C., Olariu A., Besliu P., 1987, Seminar of Archaeometry, 
Cluj, Romania\\
Craddock P.T., 1981, {\it Occasional Paper No.15}, eds. W.A. Oddy 
and W. Zwalf,\\
\hspace*{1cm}British Museum\\
Hanson V., 1978, Museum Objects in {\it X-ray Spectrometry}, ed. H.K. 
Herglotz,\\
\hspace*{1cm}L. S. Birks, 440-444\\
Roth V., 1908, Zur Glockenfunde in {\it Korrespondentzblatt des 
Vereinis f\"{u}r Siebenb\"{u}rgische\\
\hspace*{1cm}Landeskunde}, XXX, no. 9, 107-109\\
Russel M. D., Ovenden R. J., 1990, Bell founding in Winchester in
tenth to\\
\hspace*{1cm}thirteen centuries, in {\it Winchester Studies, Object and
economy in medieval\\
\hspace*{1cm}Winchester}, vol.VII, Oxford \\
\newpage
{\bf Table 1}\\

\begin{tabular}{ll}
\hline
Sample & \\
\hline
\hline
Ia      & Hearth level I\\
Ib      & Hearth level I\\
Ic      & Hearth level I\\
Id      & Hearth level I\\
Ie      & Hearth level I\\
IIa-s   & Hearth level II surface\\
IIb-s   & Hearth level II surface\\
IIc-s   & Hearth level II surface\\
IIa     & Hearth level II deepness\\
IIb     & Hearth level II deepness\\
IIc     & Hearth level II deepness\\
Za      & Slag layer\\
Zb      & Slag layer\\
Zc      & Slag layer\\
Zd      & Slag layer\\
C1      & Metallic cake, found near Hearth level I\\
C2      & Ash\\
B1      & Bell, medieval, Sibiu region\\
B2      & Bell, medieval, Sibiu region\\
B3      & Bell, medieval, Sibiu region\\
B4      & Bell, XIX century, Sibiu region\\
\hline
\end{tabular}
\newpage
{\bf Table 2}\\

\begin{tabular}{lccccccccccccccc}
\hline
Sample & Cu   & Sn   & Fe   & Au  & Ag    & Sb   & Al   & As   & Ca & K    &
 Mg    & Mn   & Na   & Sc   & Pb   \\
\hline
\hline
Ia     & 43\% &13.9\%& -    & 49   &1360  & 9050 & -    & 4490 & -  & -    &
 -     & $<$30& 98   & $<7$ & 2.7\% \\   
Ib     &70.3\%&14.5\%& -    & 48   & 1180 &1.03\%& -    & 4800 & -  & -    &
 -     &$<$40 & 200  & $<$8 & 4.4\% \\
Ic     &59.6\%&13\%  & 6.9\%& 46   & 1200 & 9000 & 970  & 4500 & -  & -    &
 -     & $<$30& 130  & $<$6 & 5.5\% \\
Id     & 69.9\%&17.6\%&8.1\% & 55   & 710  &1.22\%& 7280 & 5230 & -   & -    &
 -     &$<$70 & 930  & $<$20& 6.4\% \\
IIas     &4040  & -    &6.01\%&$<$0.5&$<$400&$<$800&14.9\%&$<$200&2\% &2.15\%&
 8\%   &5130  &1.08\%& 25   & -    \\  
IIbs     &4600  & -    &4.78\%&$<$0.4&$<$400&$<$300&13.5\%&$<200$&1\% &2.8\% &
4\%    & 448  &1.1\% & 23   & -    \\   
IIcs     &12.9\%&2.15\%&5.65\%& 3    & 310  & 1220 &1.64\%&400   & -  &0.2\% &
 -     & 102  &0.18\%& $<$7 &3.3\% \\
IIa     &29.6\%&9.78\%&11\%  & 32   & 610  & 7780 & 4.4\%& 2900 &0.8\%&0.9\% &
 -     &278   &0.32\%&11    &3\%  \\
IIb     &21.2\%&5.02\%&3.3\% & 15   & 910  & 4260 &7.2\% & 1500 & -   & 1.3\%&
 -     & 553  &0.64\%& 9    &1.6\% \\
IIc     &7.37\%&5.0\% & 1.2\%&0.7   & -    & 40   & 13\% & 220  &1.2\%& 1.9\%&
 -     & 443  &0.82\%&88    &0.6\% \\
Za     &32.4\%&5.67\%&8.5\% &19    & 490  & 4480 & 5\%   & 2390&0.7\%&0.85\%&
-      & 973  &0.70\%&13    &4.0\% \\
Zb     &25\%  &3.22\%&6.4\% &13    & 200  & 3690 & 4.8\% &  1370&0.7\%&2\%  &
 -     & 748  &1.03\%&16    &2.2\% \\
Zc     &36.4\%&11.3\%&10\%  &6     & 1040 & 9770 & 2.2\% & 3680 &0.4\%&0.87\%&
-      &660   &1.07\%&6     &5.6\% \\
C1     &16.2\%& -    & -    & 1460 & 1.4\%& 1.1\%& -     & 2500 & -   &500   &
-      &$<$2  &980   & -    & 22\% \\
C2     &7210  &6.8\% & 7.3\%&0.3   &$<$300& 33   & -     &$<$80 & -   &2.5\% &
-      & 5270 &1.46\%& 21   &0.09\% \\
\hline
\end{tabular}
\newpage
{\bf Table 3}\\

\begin{tabular}{lccccccccccc}
\hline
Sample & Cu  & Sn & Fe  &Au& Ag  & Sb   & Al  & As & Mn& Na   & Zn   \\
\hline
\hline
B1     & 74\%&19\%&5000&46 &1580 &2.1\% &3500 &7780& 60& 880  & 1000\\
B2     & 71\%&18\%&9000&42 &1530 &2.5\% &2400 &7010& 25& 400  & 500\\
B3     & 64\%&16\%&2500&85 &1950 &1.6\% &850  &8700& 30& 750  & 400\\
B4     & 72\%&14\%&5900&3  & 300 &0.58\%&4.6\%&1280& 90& 1400 & 1.7\%\\
\hline
\end{tabular}
\newpage

{\bf Table 4}\\ 

\begin{tabular}{lcccc}
\hline
Coin              & Code       & Provenance        & Ag, \%   & Cu, \%\\
\hline
\hline
Matei Corvin Denar& T.1285/4641& Sibiu              & 69.2     & 30.8   \\
Transylvania Denar& T.1285/2455& Sibiu              &35.8      & 64.2   \\
XV century Denar  &            & Saliste, near Sibiu&94.5      &  5.5   \\
Leopold I piece   & T.1285/3762&                    &57.0      & 43.0 \\
\hline
\end{tabular}
\newpage
{\bf Table 5}\\

\begin{tabular}{lccccccccccc}
\hline
Sample            & Cu     & Sn    & Fe   &Au& Ag  & Sb  & As & Pb    & Zn\\
\hline
\hline
Whitechapel Bell  & 67.87\%&25.67\%&0.06\%&200&1700 &1100&2600 &2.90\%&1.3\%\\
Christ Church Bell& 76.77\%&21.08\%&0.19\%& 0 &1100 &100 &3100 
&0.84\%&0.70\%\\
\hline
\end{tabular}
\newpage

FIGURE CAPTIONS

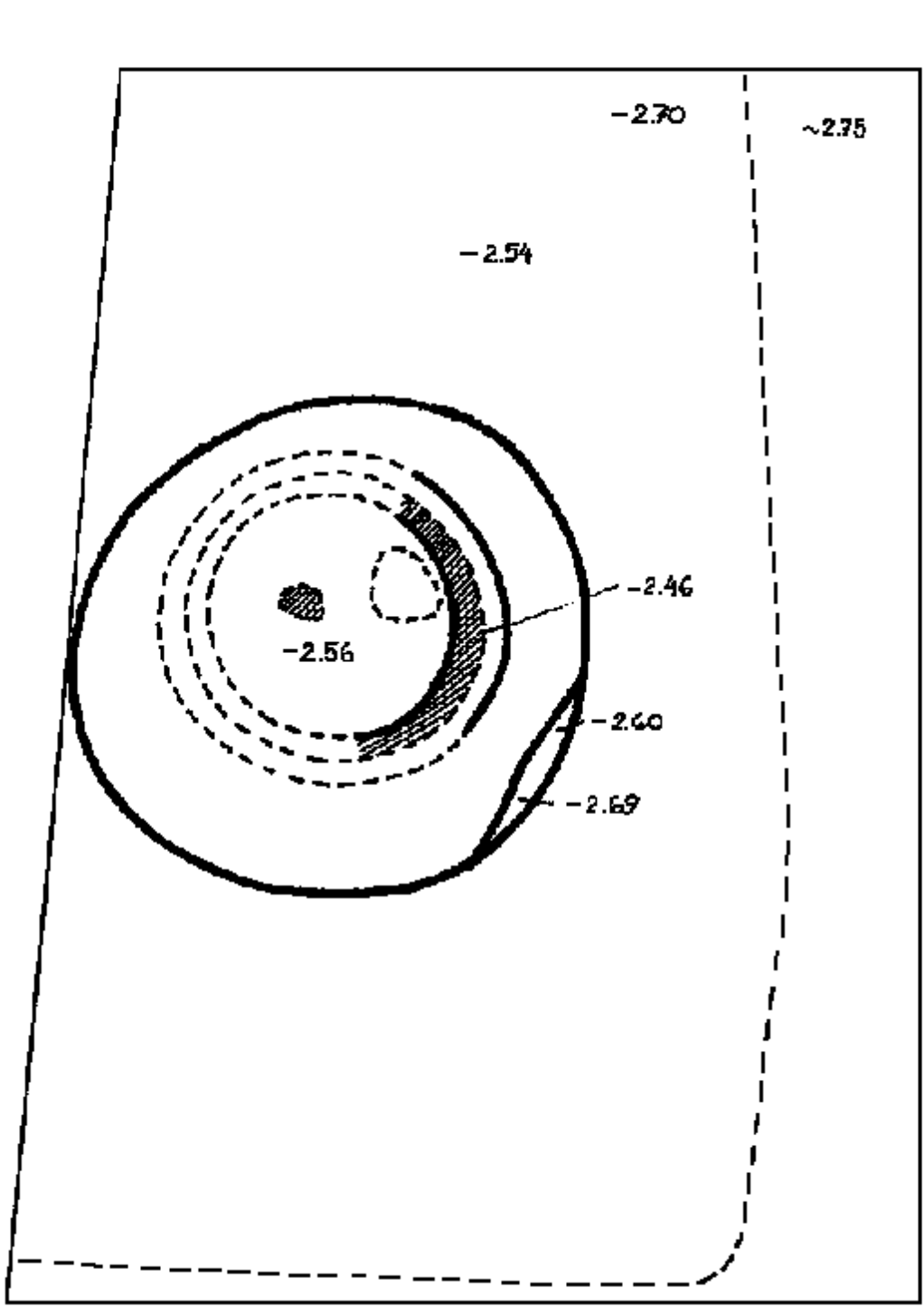
Figure 1. Schematic drawing of the Hearth level I of the bell casting pit from
Old Town Hall, Sibiu, Romania\\

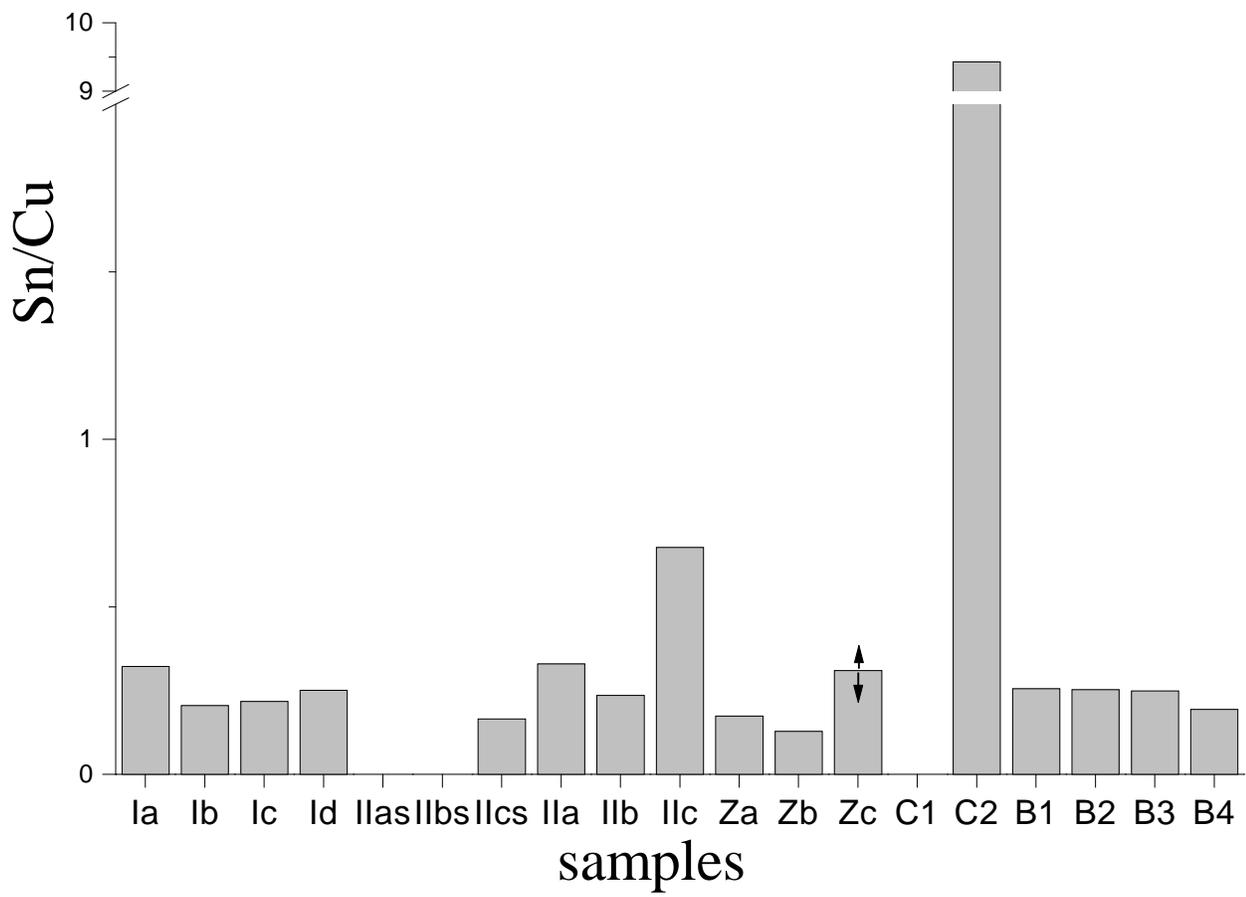
Figure 2. Ratio of the Sn and Cu concentrations for the two hearths, slag and
Transylvanian bells.\\

Figure 3. Ratio of the Fe and Cu concentrations for the two hearths, slag and
Transylvanian bells.\\

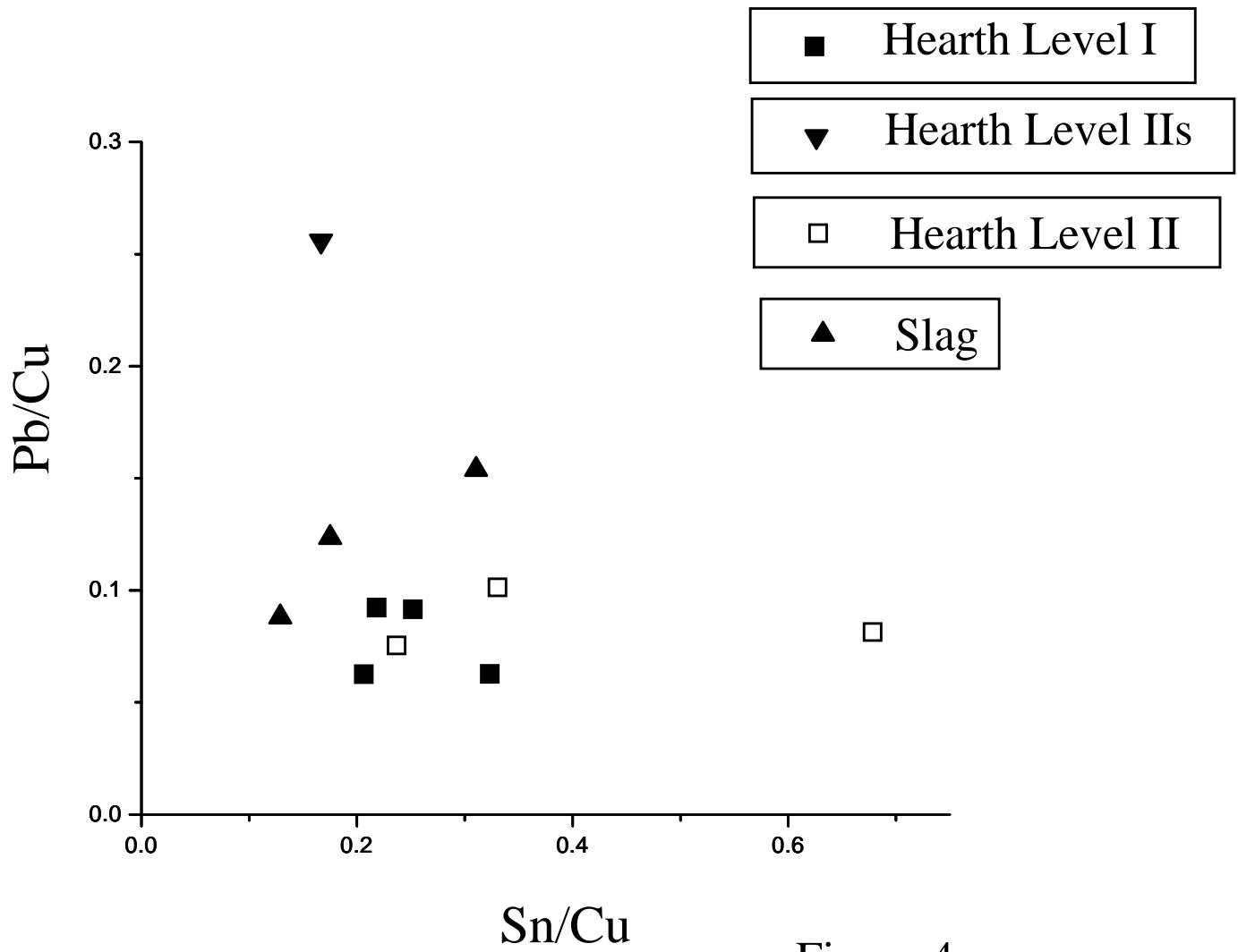
Figure 4. Diagram of Pb/Cu concentration ratio versus Sn/Cu concentration ratio
for the two hearths and slag.\\

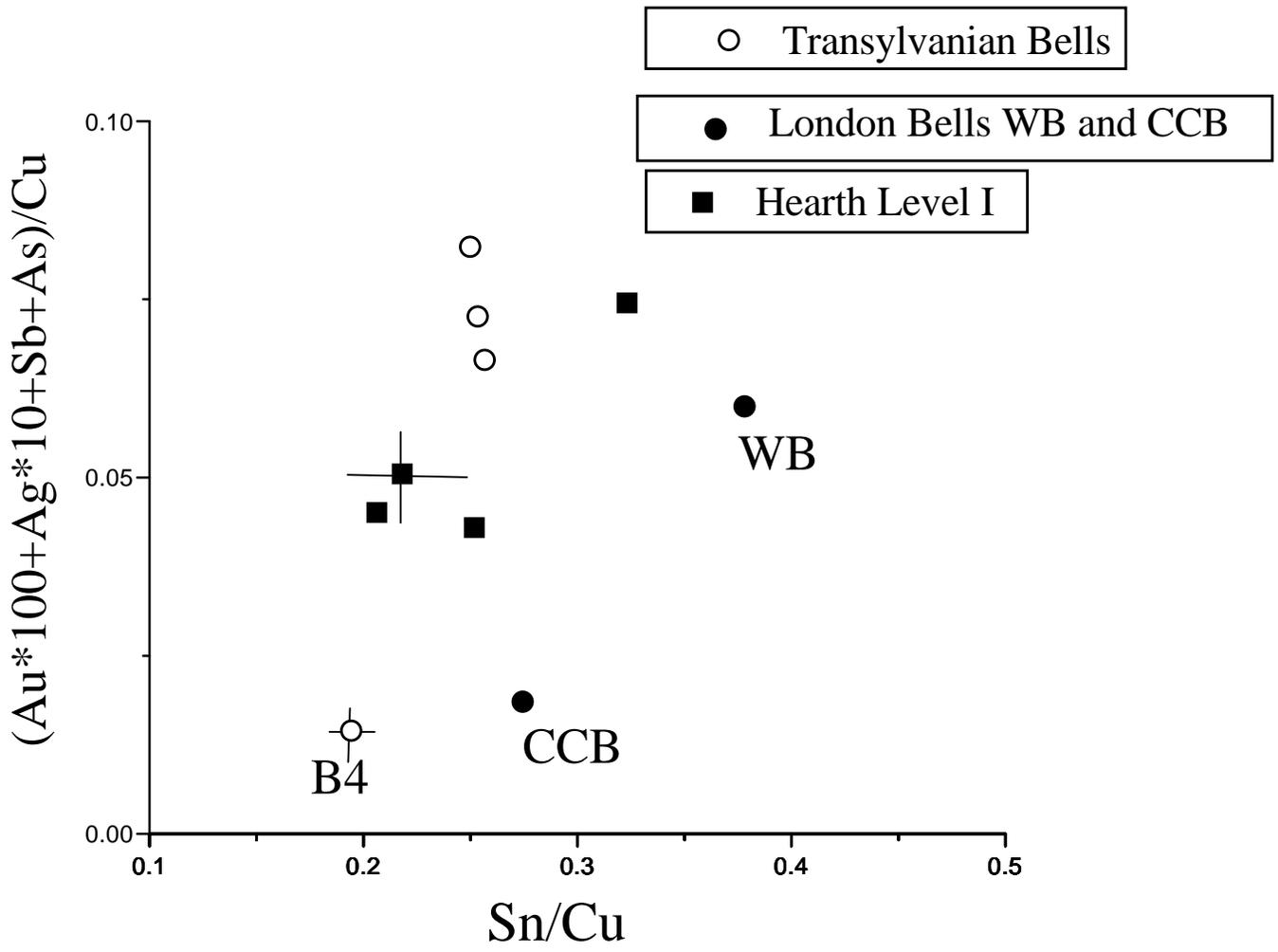
Figure 5. Diagram of composed concentration ratio (Au*100+Ag*10+Sb+As)/Cu versus
Sn/Cu concentration ratio for the Hearth level I, Transylvanian bells and 
Whitechapel bell (WB) and Christ Church bell (CCB)\\

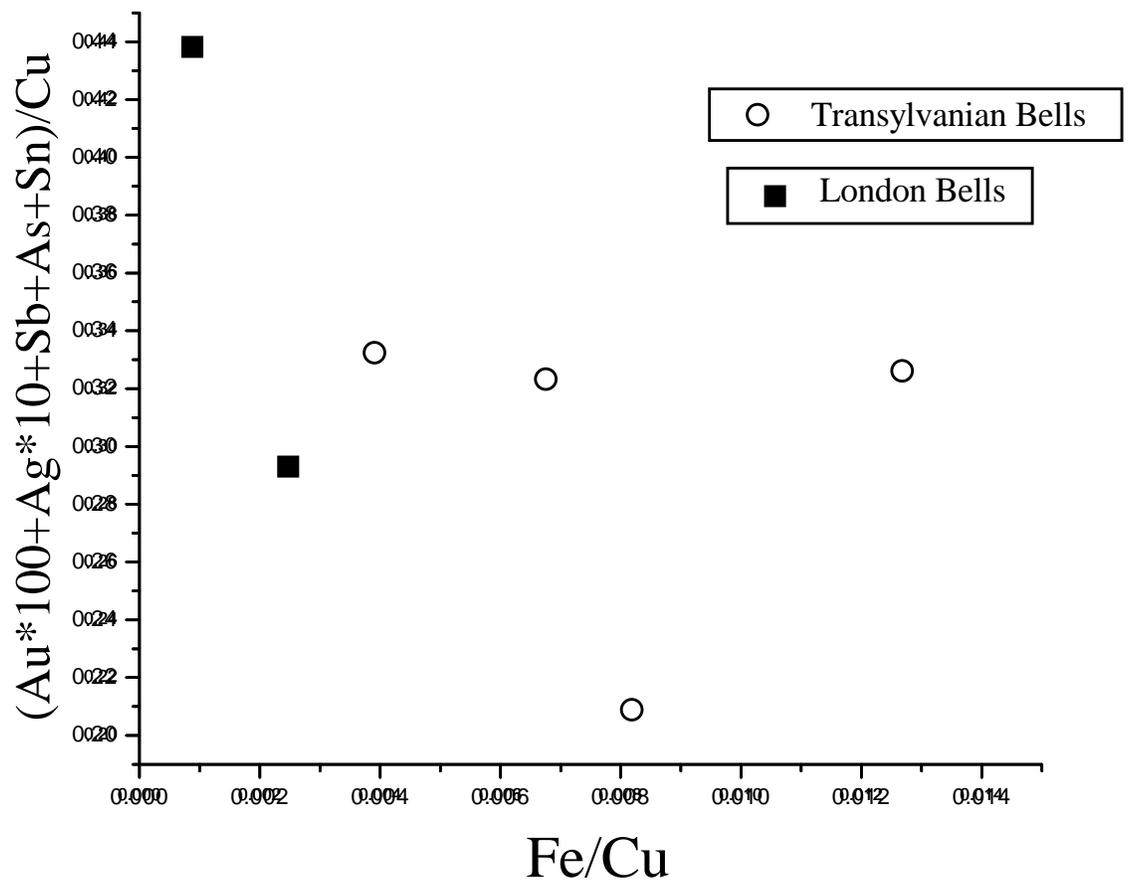
Figure 6. Diagram of composed concentration ratio (Au*100+Ag*10+Sb+As+Sn)/Cu 
versus Fe/Cu concentration ratio for Transylvanian bells and Whitechapel bell 
(WB) and Christ Church bell (CCB)\\

\newpage
TABLE CAPTIONS\\

Table 1. List of analysed samples.\\

Table 2. Concentration of elements in hearths levels I and II, slag, ash and
cake, bye NAA and XRF. The concentrations are expressed in parts per million 
(ppm), and when the
concentration is greater than 10000 ppm it is given in percents (\%).\\

Table 3. Concentration of elements in Transylvanian bells, bye NAA.
The concentrations are expressed in parts per million (ppm), and when the
concentration is greater than 10000 ppm it is given in percents (\%).\\

Table 4.  Composition of some coins from Sibiu region by gamma transmission 
(Besliu 1987)\\

Table 5. Concentration of elements for Whitechapel bell and Christ Church bell,
bye XRF (Hanson 1978).
The concentrations are expressed in parts per million (ppm), and when the
concentration is greater than 10000 ppm it is given in percents (\%).\\

\end{document}